\documentclass[12pt]{article}

\usepackage[utf8]{inputenc}
\usepackage[a4paper,top=2cm,bottom=2cm,left=2cm,right=2cm,marginparwidth=1.78cm]{geometry}
\usepackage{graphicx}
\usepackage{float}
\usepackage{bm,bbm}
\usepackage{pdflscape} 
\usepackage{siunitx}
\usepackage{booktabs} 
\usepackage{subcaption} 
\usepackage{longtable}
\usepackage{booktabs} 
\usepackage{multirow}
\usepackage{mathtools}

\usepackage{amsmath}
\usepackage{amsfonts,dsfont,amssymb}
\usepackage{xcolor}
\usepackage{listings} 
\usepackage{color} %
\usepackage{ulem}
\usepackage{hyperref}
\usepackage{times}
\usepackage{lineno}
\usepackage{natbib}

\usepackage{newtxtext,newtxmath} 
\usepackage[version=4]{mhchem} 

\usepackage{algorithm}
\usepackage{algpseudocode}
\usepackage{tikz}
\usetikzlibrary{shapes.geometric, arrows.meta, positioning}

\tikzstyle{startstop} = [rectangle, rounded corners, minimum width=3cm, minimum height=1cm, text centered, draw=black, fill=black!50]
\tikzstyle{process} = [rectangle, minimum width=5cm, minimum height=1cm, text centered, draw=black, fill=black!0]
\tikzstyle{arrow} = [thick,->,>=stealth]

\newtheorem{theo}{Theorem}[section]
\newtheorem{defi}[theo]{Definition}
\newtheorem{remark}[theo]{Remark}

\newtheorem{prop}[theo]{Proposition}

\newcommand{\dem}{{ Proof.}\ }
\begin{document}

\title{\Large Stochastic models with power-law trends: An analysis of Bitcoin price dynamics}

\author{\normalsize
	{ Alex Bomfim$^1$, Marta Calvache$^2$, Felipe Quintino$^3$\thanks{Corresponding author: felipes.quintino2@gmail.com (F. Quintino) },}\\
	{\small $^{1}$ Statistics Section of the Electoral Court of Ceará, Fortaleza-CE, Brazil}\\[-0.15cm]
{\small $^{2}$ Department of Mathematics, University of
 Sergipe, Aracaju-SE, Brazil}\\[-0.15cm]
 {\small $^{3}$ Department of Statistics, University of Brasilia, Brasilia-DF, Brazil}
}

\date{}
\maketitle
\begin{abstract}
Empirical studies have identified power-law scaling and long-memory dynamics in Bitcoin prices. To capture these features, we propose stochastic differential equations (SDEs) with a power-law trend (PLT) for modeling long-term price behavior. As a starting point, we introduce a PLT driven by a Wiener process, where the log-price follows a trend linear in log time. We explicitly derive the SDE solution, provide estimation procedures, and analyze its asymptotic properties. We extend the framework to more general driving noises. 
A Lévy-driven PLT incorporates jump behavior, whereas fractional Brownian motion captures long-range dependence. The model is extended with a power-law volatility. We present a Monte Carlo simulation study that explores the estimator’s applicability under jump-diffusion, Variance Gamma, $\alpha$-stable, and fBm noise, and under daily, weekly, and monthly sampling. To highlight our framework's applicability, we model daily Bitcoin prices, compare the PLT model with geometric Brownian motion, and select the driving noise using out-of-sample Monte Carlo forecasting.


\end{abstract}


\section{Introduction}

Bitcoin represents a digital currency that transcends traditional financial markets. Introduced in \cite{nakamoto2008bitcoin}, Bitcoin is based on blockchain technology, a digital ledger in which all transactions are securely and transparently recorded.
Transactions are carried out in a decentralized manner, not relying on intermediaries such as banks or governments, because values are transferred directly between users via digital wallets. Transactions are verified and recorded on the blockchain through a process called mining. In this process, mathematical computations are performed to validate each transaction and add new blocks to the chain sequentially. This mechanism ensures both the security and integrity of the data.
Unlike traditional currencies, Bitcoin operates continuously (24 hours a day, seven days a week) and its supply is permanently capped at 21 million units. This unique structure creates high liquidity, significant volatility, and inherent scarcity, making Bitcoin an asset that attracts the attention of both investors and researchers (cf. \cite{antonopoulos2014mastering, antonopoulos2023mastering}).

Over the past fifteen years, academic interest in bitcoin-related research topics has grown in the literature.
Several authors have investigated the presence of long memory in cryptocurrency markets, particularly in Bitcoin, suggesting that current prices may be influenced by past market movements
(\cite{bariviera2017inefficiency, wei2018liquidity, garnier2019chaos, wu2020long, li2025machine}).
Researchers have also examined the dynamics of market dependence between Bitcoin and altcoins, and between Bitcoin and green financial assets (\cite{ciaian2018virtual, yaya2022modelling, huang2023time}).
A comprehensive understanding of the multiscale behavior of Bitcoin’s price and volatility is essential for formulating robust investment strategies, a topic that has also been addressed in recent literature (\cite{bariviera2017inefficiency,garnier2019chaos,sensoy2019inefficiency, stavroyiannis2019high, dimpfl2020bitcoin, carvalho2025multiscale}).
Special attention has been devoted to Bitcoin’s market inefficiency in recent research (\cite{urquhart2016inefficiency, bariviera2017inefficiency, nadarajah2017inefficiency, sensoy2019inefficiency}). 
Finally, considerable attention has been devoted to power-law properties, particularly in analyses of Bitcoin’s price fluctuations, bubbles, and crashes (\cite{beguvsic2018scaling, wheatley2019bitcoin, shu20212021, sornette2025transaction}).
We refer the reader to \cite{MeredizSol2019, Rejeb2023} for a comprehensive review of the Bitcoin literature.

The goal of this paper is to propose a new class of processes for the logarithm of Bitcoin prices, with power-law deterministic trends on a log-time scale. This class of stochastic models is described as SDEs, providing a flexible framework for modeling non-stationary stochastic dynamics.
We consider that the Bitcoin price (in dollars) at time $ t $, is given by the model
\begin{equation}\label{eq:model_intro}
P(t) = P_0 t^\beta e^{\sigma N(t)},
\end{equation}
where $\boldsymbol{\theta}=(\beta, \sigma)^\top$ are unknown parameters, $N(t)$ is a stochastic process and $P_0$ is an independent initial value.

We propose the PLT model driven by a Wiener process, which was constructed based on the power-law regression analysis 
of Bitcoin, and we derive least squares estimators (LSE) and their exact distribution under Brownian noise. Our framework extends the PLT by replacing the noise by Lévy processes (jumps and heavy tails) and by fBm (long memory) to identify which characteristics most strongly influence the dynamics.
We propose a filtered LSE as an estimator alternative, which coincides with the standard LSE under Brownian noise but achieves higher precision in the presence of jumps and heavy-tailed noise.
We also introduce a power-law volatility structure to capture the decay of Bitcoin volatility over time.
The PLT models are compared with the geometric Brownian motion in forecasting daily Bitcoin prices.

Models based on SDEs have been widely used to analyze financial data, especially those models with memory effects, which play a significant role. 
A properly specified drift term enables the model to capture long-horizon dynamics, allowing for more reliable predictions of future trends in the time series under study. Accurate parameter estimation in SDEs is essential to ensure that the mathematical model aligns closely with real-world data. Moreover, the choice of stochastic noise in the SDE can significantly affect the performance of likelihood-based estimation methods.

The rest of the article is organized as follows. In Section~\ref{sec:3}, we introduce the PLT SDE, the LSE and its properties under Brownian, L\'evy, and fBm noises, and the extension with power-law volatility. In Section~\ref{sec:simulation}, we report Monte Carlo results. In Section~\ref{sec:application_real_data}, we apply the framework to daily Bitcoin prices. Finally, in Section~\ref{sec:6}, we present the concluding remarks.

\section{Theoretical framework for Bitcoin price} \label{sec:3}

In this section, we derive the PLT SDE, its LSE, and their extension to Lévy and fBm noises, and to a power-law volatility.

\subsection{A SDE with power-law trend}

Define the process $\mathbf{P} = \{P(t);\ t > 0\}$ as
\begin{equation}\label{eq:power_law_solution}
    P(t) = P_0\, t^{\beta} e^{\sigma B(t)}, \quad t > 0,
\end{equation}
where $P_0 > 0$ is the initial scale, $\beta \in \mathbb{R}$, $\sigma > 0$,
and $\mathbf{B} = \{B(t);\ t \geq 0\}$ is a Wiener process (standard Brownian motion).
Taking logarithms gives
\begin{equation}\label{eq:linear_reg}
    \ln P(t) = \ln P_0 + \beta \ln t + \sigma B(t), \quad t > 0,
\end{equation}
which has the structure of a linear regression\footnote{Without any connection to SDEs or Wiener processes, informal references have reported log–log behavior based on regression analysis. See, for example, \url{https://medium.com/quantodian-publications/bitcoins-natural-long-term-power-law-corridor-of-growth-649d0e9b3c94} (accessed on 02-21-2026).} in $\ln t$.
Denote the partial derivatives of $f(t,x) = t^{\beta}e^{\sigma x}$ 
as $\partial_t f(t,x)$, $\partial_x f(t,x)$, and 
$\partial^2_{xx} f(t,x) = \partial_x\!\left(\partial_x f(t,x)\right)$.
Applying the It\^{o} formula (see, e.g., (2.26) in 
\cite{mikosch1998elementary}),
$$
df(t, B(t)) = \left[\partial_t f(t,B(t)) 
+ \frac{1}{2}\partial^2_{xx} f(t,B(t))\right]dt 
+ \partial_x f(t,B(t))\,dB(t),
$$
we obtain that the power-law trend (PLT) SDE $P(t) = P_0 f(t, B(t))$ satisfies the homogeneous 
linear SDE
\begin{equation}\label{eq:sde_power_law}
    dP(t) = \left[\beta t^{-1} + \frac{1}{2}\sigma^2\right]P(t)\,dt 
    + \sigma P(t)\,dB(t), \quad t > 0.
\end{equation}

We are interested in estimating the parameter $\boldsymbol{\theta}=(\beta, \sigma)$ of the SDE with power-law trend \eqref{eq:sde_power_law}. Our proposed estimators are based on an appropriate discretization of the SDE. 

We need to fix some notation. Consider that the process \eqref{eq:power_law_solution} is observed at times $0<t_0<t_1<\cdots<t_n=T$, where $t_k=t_0+kh$ with $h=(T-t_0)/n$. 
The increment of a function $f$ over the subinterval $[t_i, t_{i+1}]$ is denoted by $\Delta_k f = f(t_{k+1})-f(t_k)$, and the interval increment and log-increment by $\Delta_k=t_{k+1}-t_k$ and $r_k=\ln(t_{k+1}/t_k)$, $k=0,\ldots,n-1$. We write $P_{t_k}=P(t_k)$.

\begin{figure}[H]
\centering
\includegraphics[width=1.0\linewidth]{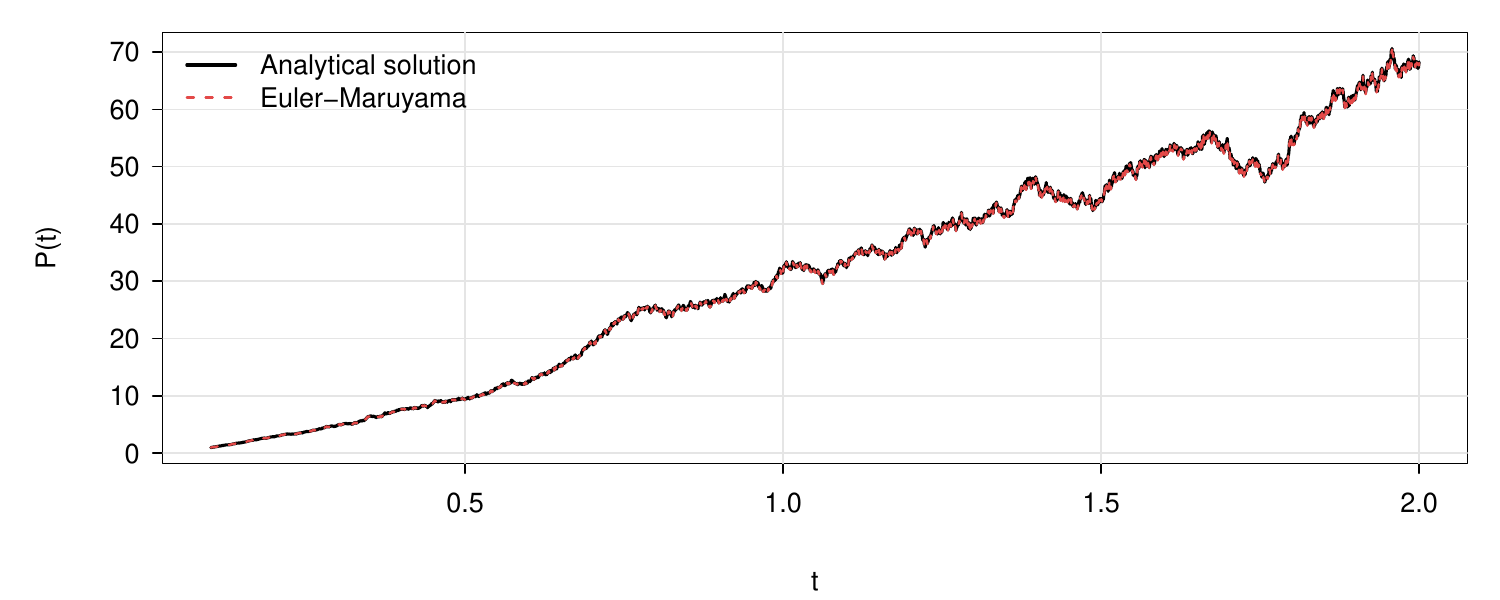}
\caption{Comparison between the Euler approximation of the PLT SDE and its analytical solution ($\beta=1.5$, $\sigma=0.3$, $P_0=1$, $t\in[0.1,2]$, $n=10,000$).}
\label{fig:euler}
\end{figure}

\begin{figure}[H]
\centering
\includegraphics[width=1.0\linewidth]{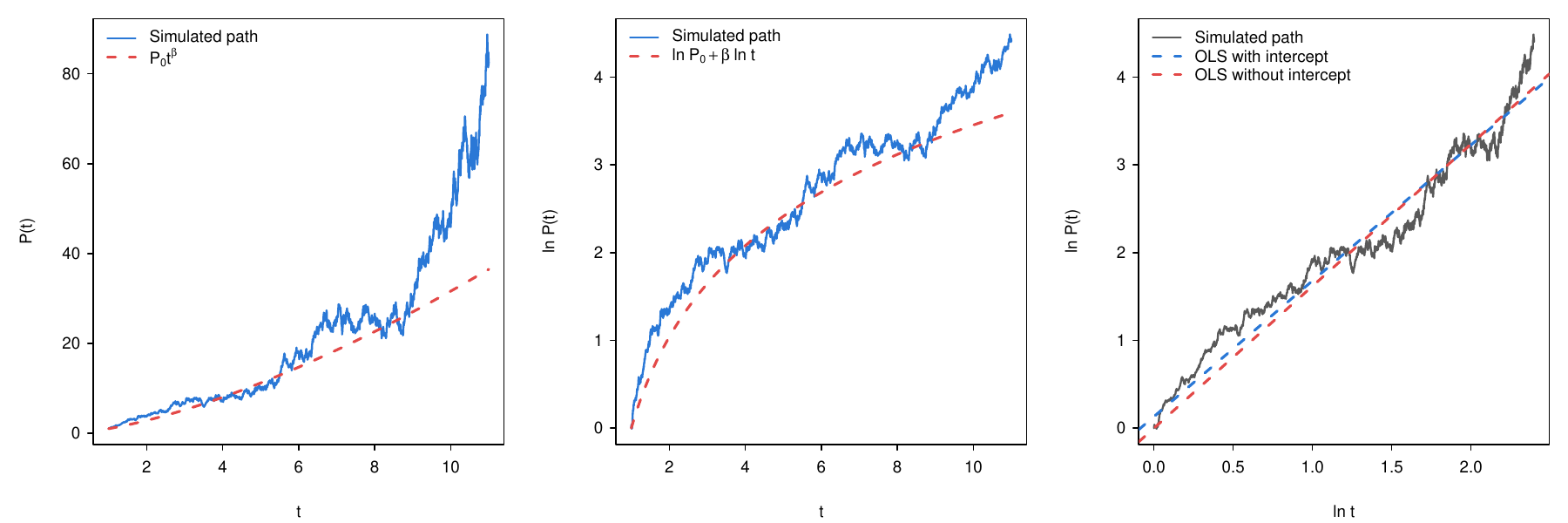}
\caption{Path of the PLT SDE $P(t)$ (left) with a deterministic trend shown as a red dashed line, the 
$\ln P(t)$ plot (center), and the log–log plot (right) with regressions fitted with (blue dashed line) and without (red dashed line) an intercept ($\beta=1.5$, $\sigma=0.3$, $P_0=1$, $t\in[1,11]$, $h=1/365$).}
\label{fig:path}
\end{figure}

Figure \ref{fig:euler} shows a path comparison between the Euler approximation of the PLT SDE and its analytical solution. The Euler approximation can accurately capture the analytical solution, allowing us to estimate the model parameters using LSE.
Figure \ref{fig:path} shows that the process plotted on a log–log scale follows the trend of a linear regression model, as expected from its construction in \eqref{eq:power_law_solution}--\eqref{eq:sde_power_law}.

By Itô's formula, 
\begin{equation}\label{eq:log-plt}
    d\ln P(t)=\beta t^{-1}dt+\sigma\, dB(t).
\end{equation}
Since $\int_{t_k}^{t_{k+1}}t^{-1}dt=r_k$, the Euler scheme of this equation is
\begin{equation}\label{eq:increments}
\Delta_k \ln P=\beta r_k+\sigma\,\Delta_k B,\qquad k=0,\ldots,n-1.
\end{equation}
Minimizing $$\sum_{k=0}^{n-1}(\Delta_k\ln P-\beta r_k)^2$$ yields the LSE $(\hat{\beta}_n, \hat{\sigma}^2_n)^\top$,
\begin{equation}\label{eq:lse}
\hat{\beta}_n = \frac{\sum_{k=0}^{n-1} r_k\,\Delta_k\ln P}{\sum_{k=0}^{n-1} r_k^2},
\qquad
\hat{\sigma}^2_n = \frac{1}{n}\sum_{k=0}^{n-1}\frac{\left(\Delta_k\ln P - \hat{\beta}_n r_k\right)^2}{\Delta_k}.
\end{equation}

The following result shows $ \hat{\bm\beta}_n $ has a normal distribution, is unbiased, but is not consistent.

\begin{prop}\label{prop:lse}
Denote $S_n=\sum_{k=0}^{n-1}r_k^2$. Then, the estimators \eqref{eq:lse} of the PLT model \eqref{eq:sde_power_law} satisfies:
\begin{itemize}
\item[(a)] $\hat\beta_n\sim N(\beta,\sigma^2h/S_n)$ and $n\hat\sigma_n^2/\sigma^2\sim\chi^2_{n-1}$, independent of $\hat\beta_n$.
\item[(b)] As $n\to\infty$, $h/S_n\to t_0T/(T-t_0)$ and $\hat\sigma_n^2\rightarrow\sigma^2$ in mean square. 
\end{itemize}
\end{prop}

\dem Denote $Z_k$ i.i.d. $N(0,1)$. By the Euler scheme \eqref{eq:increments}, $$\Delta_k\ln P=\beta r_k+\sigma\sqrt{h}Z_k,$$
is a Gaussian linear model without intercept, and (a) follows from Cochran's theorem.

For (b), using a Taylor expansion, we have $$r_k=h/t_k+O(h^2),$$
which gives $$S_n/h\to\int_{t_0}^{T}t^{-2}dt=1/t_0-1/T,$$ and (a) gives $\mathbb{E}(\hat\sigma^2_n-\sigma^2)^2=\sigma^4(2n-1)/n^2\to0,$ as $n\rightarrow\infty$. $\square$

\begin{remark}\label{rem:general}
\begin{enumerate}
\item It follows from Proposition~\ref{prop:lse} that $\hat\beta_n$ is unbiased but not consistent, because its RMSE converges to $\sigma\{t_0T/(T-t_0)\}^{1/2}$, which tends to $\sigma\sqrt{t_0}$ as $T\to\infty$.
    \item  The exact confidence interval for $\beta$ is
    $$\hat\beta_n\pm t_{n-1;0.975}\{n\hat\sigma^2_nh/((n-1)S_n)\}^{1/2}.$$
    Moreover, since the increments \eqref{eq:increments} are independent and Gaussian, $\hat\beta_n$ attains the Cramér--Rao lower bound. 

    \item The lack of consistency is not due to the Gaussian noise. If $B$ in \eqref{eq:increments} is replaced by any process $N$ with c\`adl\`ag paths, summation by parts gives, path by path,
\begin{equation}\label{eq:limit_beta}
\hat\beta_n\to\beta+\sigma\left(\frac{1}{t_0}-\frac{1}{T}\right)^{-1}\int_{t_0}^{T}t^{-1}\,dN(t),\qquad n\to\infty,
\end{equation}
a non-degenerate limit. 

\item In Section \ref{sec:plv}, we develop a consistent estimator by extending the PLT SDE framework to a power-law stochastic volatility model.
\end{enumerate}

\end{remark}

\subsection{The SDE with power-law trend driven by general random noises}\label{sec:general}

In this subsection, we generalize the PLT SDE by incorporating jumps through a Lévy process or long-term memory via fBm. For the Lévy case, adaptations to the LSE are presented, while noise parameter estimation is addressed in both settings.

In deriving LSE \eqref{eq:lse}, we used only the increments of $\ln P$, not the noise law. This allows us to replace $\mathbf{B}$ with noise that better fits Bitcoin behavior. Given a driving process $\mathbf{N}=\{N(t);~t\geq0\}$, we define the PLT model through the log-price,
\begin{equation}\label{eq:general}
d\ln P(t)=\beta t^{-1}dt+\sigma\, dN(t),\qquad t>t_0,
\end{equation}
Its solution is
$$P(t)=P(t_0)(t/t_0)^{\beta}e^{\sigma\{N(t)-N(t_0)\}}, ~~t>t_0.$$ 
For the classical PLT process, set the random noise to be the Wiener process $\mathbf{N}=\mathbf{B}$.

For the reader's convenience, the definition of the Lévy process and the fBm are presented below.
	\begin{defi}
		We say that a stochastic process $\textbf{L}=\{L(t); t\geq 0\}$ is a \textsf{Lévy process} if
  
			a) $L(0)=0$ almost surely (a.s.);

			b) $\textbf{L}$ has independent increments, i.e., $L(t_0), L(t_1)-L(t_0), \cdots L(t_n)-L(t_{n-1})$ are independent random variables for every $0<t_0<t_1<\cdots<t_{n-1}<t_n$ and for all positive integer $n$;

			c) $\textbf{L}$ has stationary increments, i.e., for all $t\geq0$ and all $h>0$, $L(t+h)-L(t)$ has the same distribution as $L(h)$;

			d) $\textbf{L}$ is càdlàg (right continuous with left limits) paths and is stochastically continuous,
			\begin{displaymath}
				\lim\limits_{t\rightarrow s} \mathbb{P}\left(|L(t) - L(s)|>\delta \right) = 0,\,\,\forall \delta>0\,\,\mbox{and}\,\,\forall s >0. 
			\end{displaymath} 
				
		\end{defi}
We consider Lévy processes $\mathbf{L}$ with $\mathbb{E}L(1)=0$. Wiener process, Poisson process, Compound Poisson process,  variance gamma (VG), and symmetric $\alpha$-stable are typical examples of Lévy processes.

\begin{defi}
We say that $\mathbf{B}^H=\{B^H(t); t\geq0\}$  is a \textsf{fractional Brownian motion} (fBm) with Hurst parameter $H\in(0,1)$ if it is a centered Gaussian process with the covariance function
$$R_H(t,s) = \frac{1}{2} \left( s^{2H}+t^{2H}-|t-s|^{2H} \right).$$
\end{defi}

Since $\mathbb{E} \Delta_kN=0$, $\hat\beta_n$ is unbiased for all Lévy processes above and the fBm, with the same variance as in Proposition~\ref{prop:lse}. In contrast, $\hat\sigma^2_n$ estimates $\sigma^2$ only for finite-variance Lévy noises, diverging as $h\to0$ for stable noise, and $\mathbb{E}\hat\sigma^2_n\approx\sigma^2h^{2H-1}$ for fBm. The noise parameters are estimated from the residual increments 
\begin{equation}\label{eq:residual}
    \hat e_k=\Delta_k\ln P-\hat\beta_n r_k.
\end{equation}

Under Lévy noise with jumps or heavy tails, a few large increments dominate $\hat\beta_n$. Using the jump filtering technique of \cite{Mancini09} (see also \cite{quintino2026asymptotics} and the references therein), we propose the filtered LSE (FLSE) $\widetilde\beta_n$, defined as
\begin{equation}\label{eq:flse}
\widetilde\beta_n=\frac{\sum_{k=0}^{n-1} r_k\,\Delta_k\ln P \mathbbm{1}_{\left[ |\Delta_k \ln P|\leq a_n^k h^b \right]}}{\sum_{k=0}^{n-1}r_k^2\mathbbm{1}_{\left[ |\Delta_k \ln P|\leq a_n^k h^b \right]}},
\end{equation}
where $a_n^k$ is measurable with respect to the observations $\{P(t_0), P(t_1),\cdots, P(t_k)\}$ and $b\in(0,1/2)$.

\subsection{The SDE with power-law trend and power-law volatility}\label{sec:plv}

The assumption of constant $\sigma$ in \eqref{eq:log-plt} and \eqref{eq:general} limits the derivation of consistent estimators. In real-world scenarios, Bitcoin volatility decreases over time (see Section \ref{sec:application_real_data}). In this subsection, we consider the scale decay as a power of time
\begin{equation}\label{eq:plv}
d\ln P(t)=\beta t^{-1}dt+\sigma(t)\,dN(t),\qquad \sigma(t)=\sigma_0\Bigl(\frac{t}{t_0}\Bigr)^{-\gamma},\qquad t>t_0,
\end{equation}
with $\sigma_0>0$ and $\gamma\in\mathbb{R}$. Note that $\gamma=0$ recovers \eqref{eq:general}.

Consider the Wiener noise $\mathbf{N}=\mathbf{B}$. Itô's formula gives us 
$$dP(t)=\left[\beta t^{-1}+\sigma^2(t)/2\right]P(t)\,dt+\sigma(t)P(t)\,dB(t).$$
The discrete-time process corresponding to \eqref{eq:plv} is
\begin{equation}\label{eq:plv_inc}
\Delta_k\ln P=\beta r_k+\sigma_0\,v_k^{1/2}(\gamma)\,Z_k,\qquad v_k(\gamma)=\int_{t_k}^{t_{k+1}}\Bigl(\frac{s}{t_0}\Bigr)^{-2\gamma}ds,
\end{equation}
with $Z_k$ i.i.d. $N(0,1)$.

We propose a two-step estimation procedure for the parameter $\boldsymbol{\theta} = (\beta, \sigma_0, \gamma)^\top$. In the first step, given $\gamma$, denote $J_n(\gamma)=\sum_{k=0}^{n-1}\frac{r_k^2}{v_k(\gamma)}.$
The Gaussian likelihood is then maximized by the weighted LSE (WLSE)
\begin{equation}\label{eq:wlse}
\hat\beta_n(\gamma)=\frac{1}{J_n(\gamma)}\sum_{k=0}^{n-1}\frac{r_k\,\Delta_k\ln P}{v_k(\gamma)},\qquad
\hat\sigma^2_0(\gamma)=\frac{1}{n}\sum_{k=0}^{n-1}\frac{\{\Delta_k\ln P-\hat\beta_n(\gamma)r_k\}^2}{v_k(\gamma)}.
\end{equation}
In the second step, $\hat\gamma_n$ maximizes the profile log-likelihood, that is,
\begin{equation}\label{eq:wlse2}
    \hat\gamma_n \in \arg\max_{\gamma\in\mathbb{R}} -\frac12\sum_{k=0}^{n-1}\ln\left(\hat\sigma_0^2(\gamma)v_k(\gamma)\right).
\end{equation}
For simplicity of notation, we write $\hat\beta_n=\hat\beta_n(\hat\gamma_n)$ and $\hat\sigma_0^2=\hat\sigma_0^2(\hat\gamma_n)$.

We now derive the asymptotic behavior of the estimator \eqref{eq:wlse}--\eqref{eq:wlse2}. If $\gamma=0$, the new result reduces to Proposition~\ref{prop:lse}. Now, we show that if $\gamma\geq1/2$, $\hat\beta_n(\gamma)$ is a consistent estimator for $\beta$.

\begin{theo}\label{prop:plv}
Consider the PLT SDE with power-law volatility \eqref{eq:plv} and $\gamma$ being a known parameter. Then,
\begin{itemize}
\item[(a)] $\hat\beta_n(\gamma)\sim N(\beta,\sigma_0^2/J_n(\gamma))$ and $n\hat\sigma_0^2(\gamma)/\sigma_0^2\sim\chi^2_{n-1}$, independent of $\hat\beta_n(\gamma)$.
\item[(b)] $$J_n(\gamma)\to J_T(\gamma)=t_0^{-2\gamma}\int_{t_0}^{T}t^{2\gamma-2}dt, ~~\text{as}~~n\to\infty,$$
and $J_T(\gamma)/\sigma_0^2$ is the Fisher information on $\beta$ of the continuous-time process $\{\ln P(t);\ t_0\le t\le T\}$.
\item[(c)]  $\hat\beta_n(\gamma)$ is consistent if and only if $\gamma\ge1/2$.
\end{itemize}
\end{theo}
\dem As in Proposition~\ref{prop:lse}, since \eqref{eq:plv_inc} is a Gaussian linear model with known weights, this proves (a).  Note that $$r_k^2/v_k(\gamma)=h\,t_0^{-2\gamma}t_k^{2\gamma-2}\{1+O(h)\}.$$ 
The convergence in (b) arises from the definition of  $J_k(\gamma)$ and holds as $n\rightarrow\infty$. 
It follows from Girsanov's theorem that the Fisher information of the path is $$\int_{t_0}^{T}t^{-2}\sigma^{-2}(t)\,dt=J_T(\gamma)/\sigma_0^2.$$ 
Finally, (c) follows from evaluating $\int_{t_0}^{T}t^{2\gamma-2}dt$, and from the asymptotic relation 
$$r_k^2/v_k(\gamma)\sim h\,t_0^{-2\gamma}t_k^{2\gamma-2} ~~\text{as}~~ t_k\to\infty,$$
for fixed $h$, which implies
$$J_T(\gamma) \rightarrow \infty ~~\text{as}~~T\rightarrow\infty, $$
if and only if $\gamma\geq1/2$.
$\square$

\begin{remark}
For the Lévy and fBm noises, in \eqref{eq:plv} we take $\sigma(t)$ constant within each sampling interval, that is, we replace $Z_k$ with $h^{-1/2}\Delta_kN$.
\end{remark}

\section{Simulation results}\label{sec:simulation}

In this section, we present Monte Carlo simulations to assess the PLT estimates.
We analyze the PLT models driven by the Wiener process, the fBm, and the Lévy process $L(t)$, under three distinct scenarios: a Wiener process combined with a compound Poisson process (W+CPP), a VG process, and a Stable process.
For the W+CPP, we consider a finite-activity Lévy process $L(t)$ in which the jump component is a sum of $Y_j\sim N(0,2)$, which are i.i.d. random variables. 

We assess the performance of the LSE, FLSE, and noise parameter estimates using the mean absolute error (MAE) and the root mean square error (RMSE) under different noise scenarios. 

We conducted $M=1{,}000$ Monte Carlo replications, generating samples $\{X({t_0}), X({t_1}),\cdots, X({t_n})\}$ drawn from the Euler-Maruyama scheme of \eqref{eq:log-plt}, \eqref{eq:general} and \eqref{eq:plv}. We fix $\beta=1.5$, $t_0=1$ (time in years) and, within each subsection, vary $\sigma$, the noise parameter, and the sample size $n$.

The simulation study was programmed in R software 4.5.1. 

\subsection{Evaluating the estimates of the Wiener-driven PLT model}

In this subsection, we present a Monte Carlo simulation study to evaluate the parameter estimates for \eqref{eq:log-plt}. 

\begin{figure}[H]
\centering
\includegraphics[width=0.9\linewidth]{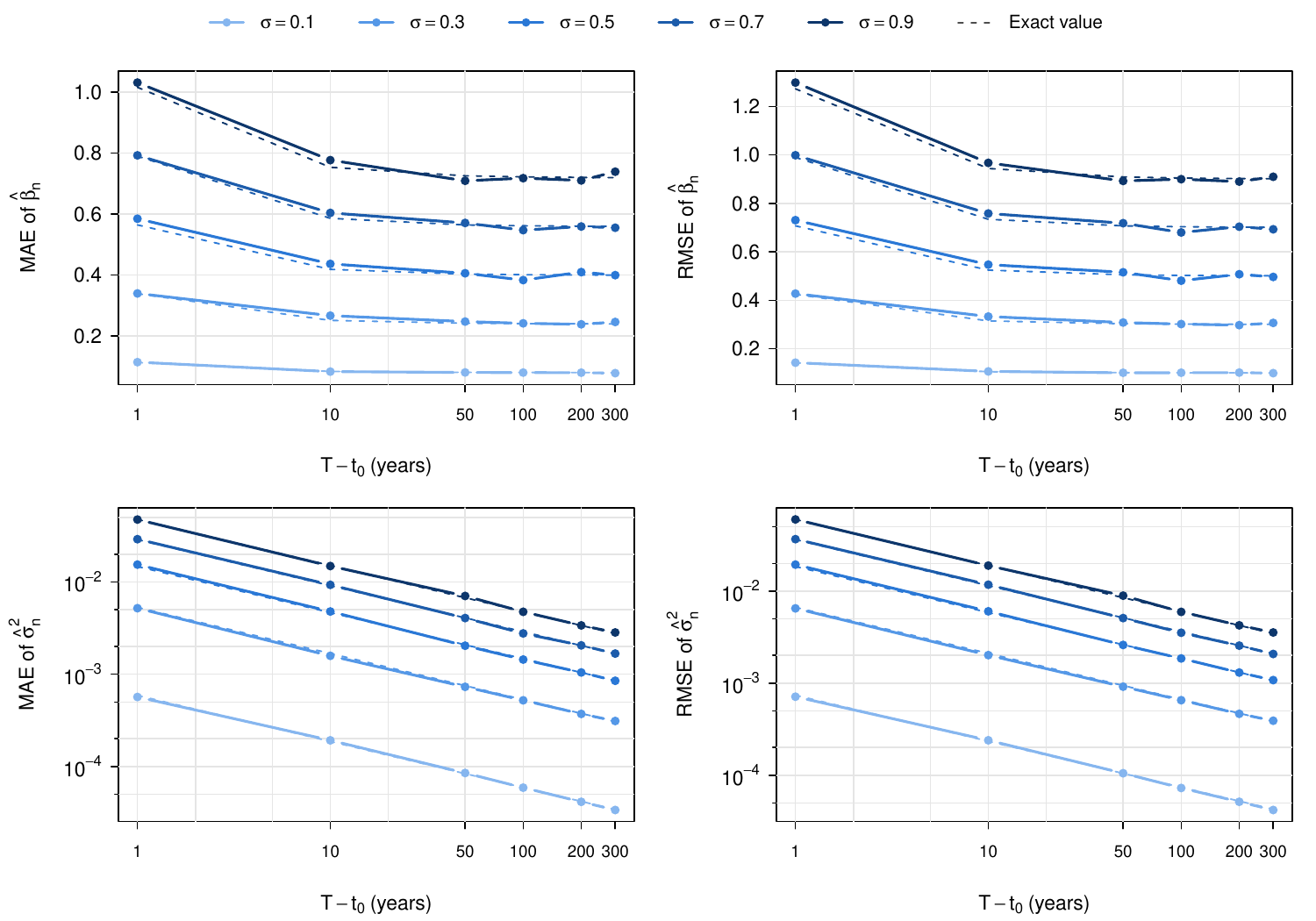}
\caption{MAE (left) and RMSE (right) for the LSE of $\bm\theta = (\beta, \sigma^2)^\top$ based on Monte Carlo simulations of the PLT model \eqref{eq:log-plt} driven by a Wiener process. Results for $\beta$ are shown on top, and $\sigma^2$ on the bottom. Dashed lines have the exact values from Remark~\ref{rem:general}.
}
\label{fig:mc1}
\end{figure}

Consider $T-t_0\in\{1,10,50,100,200,300\}$ and $\sigma\in\{0.1,0.3,0.5,0.7,0.9\}$. Figure~\ref{fig:mc1} presents MAE and RMSE for \eqref{eq:lse} using daily sampling ($h=1/365$).
The results show that the MAE and RMSE of $\hat{\sigma}_n$ decrease as $T$ increases. For $\hat{\beta}_n$, both metrics tend to stabilize around the exact values stated in Remark~\ref{rem:general}, without converging to zero as $T$ increases. Improving the estimation of $\beta$ requires a power-law volatility model (see Subsection~\ref{sec:sim_plv}).
Lower values of $\sigma$ yield improved accuracy for both estimators.

\begin{figure}[H]
\centering
\includegraphics[width=0.9\linewidth]{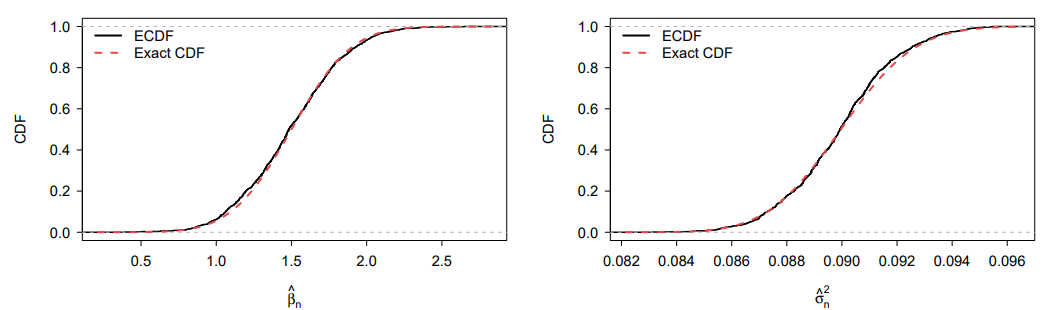}
\caption{Empirical versus exact distribution functions of the Monte Carlo estimates $\hat\beta_n$ (left) and $\hat\sigma^2_n$ (right) with fixed parameters $\beta=1.5$, $\sigma=0.3$, $T-t_0 = 10$, and $h=1/365$.}
\label{fig:mc2}
\end{figure}

To analyze the approximation laws in Proposition~\ref{prop:lse}, we fix $\sigma=0.3$ and $T-t_0=10$. Figure \ref{fig:mc2} shows that the empirical distribution functions (ECDFs) of the LSE $\hat{\beta}_n$ and $\hat{\sigma}_n^2$ align with the exact normal and scaled chi-square distributions. The sample correlation obtained was $-0.002$.

\subsection{Evaluating the estimates of the PLT model driven by fBm and Lévy process}

In this subsection, we assess performance for the LSE and FMLE in the fBm- and Lévy-driven PLT model.

Table~\ref{tab:noises} presents the results of $\hat\beta_n$, $\widetilde\beta_n$ and $\hat\sigma^2_n$, based on $M=1{,}000$ Monte Carlo simulations of the PLT model \eqref{eq:general} driven by fBm and Lévy processes. The estimation of the noise parameter $\vartheta\in\{\lambda, \nu, \alpha, H\}$ is also addressed, where $\lambda$, $\nu$, $\alpha$ and $H$ correspond to W+CPP, VG, stable, and fBm, respectively.
We fix  $\sigma=0.5$, $\lambda=10$, $\nu=0.02$, $\alpha=1.5$ and $H=0.3$ (short-range dependence) and $H=0.7$ (long-range dependence), while varying $T-t_0\in\{1,10,100,300\}$ to assess the convergence of MAE and RMSE.

Results in Table~\ref{tab:noises} show that both MAE and RMSE of  $\hat\sigma_n^2$ decrease as $T$ increases. 
For $\hat\beta_n$ and $\widetilde\beta_n$, the metrics are nearly identical for Wiener noise and fBm with $H=0.7$ (long-range dependence), with no improvement as $T$ grows. 
Considering W+CPP, VG, and stable noises, $\widetilde\beta_n$ yields lower MAE and RMSE values compared with $\hat\beta_n$. However, the results remain unsatisfactory and improve only under stochastic volatility (see Subsection \ref{sec:app_vol}). For fBm noise with $H=0.3$, $\hat\beta_n$ performs better than $\widetilde\beta_n$, yet the conclusions remain non-consistent, as in the previous cases.

The noise parameters are estimated using the mean of the residuals from the fitted model \eqref{eq:residual}. The last two columns of Table~\ref{tab:noises} report good results for VG, stable, and fBm (both short- and long-range dependence), even in cases where the power parameter $\beta$ was not well estimated. For W+CPP noise, although $\hat\lambda$ improves as $T$ increases, $T-t_0=300$ is still insufficient to yield satisfactory results.

\begin{table}[H]
\centering\small
\caption{MAE and RMSE of the LSE $\hat\beta_n$, the FLSE $\widetilde\beta_n$, the scale $\hat\sigma^2_n$ and the noise parameter $\hat\vartheta$ based on $M=1{,}000$ Monte Carlo simulations of the PLT model \eqref{eq:general} driven by fBm and Lévy processes with $\sigma=0.5$ and $h=1/365$.}
\label{tab:noises}
\begin{tabular}{llccccccccc}
\toprule
 & & \multicolumn{2}{c}{$\hat\beta_n$ } & \multicolumn{2}{c}{$\tilde\beta_n$ } & \multicolumn{2}{c}{$\hat\sigma^2_n$} & & \multicolumn{2}{c}{$\hat\vartheta$} \\
\cmidrule(lr){3-4} \cmidrule(lr){5-6} \cmidrule(lr){7-8} \cmidrule(lr){10-11}
Noise & $T-t_0$ & MAE & RMSE & MAE & RMSE & MAE & RMSE & $\vartheta$ & MAE & RMSE \\
\midrule
Wiener & 1 & 0.554 & 0.695 & 0.560 & 0.704 & 0.0148 & 0.0183 & -- & -- & -- \\
 & 10 & 0.405 & 0.504 & 0.407 & 0.505 & 0.0045 & 0.0056 &  & -- & -- \\
 & 100 & 0.397 & 0.493 & 0.401 & 0.497 & 0.0014 & 0.0018 &  & -- & -- \\
 & 300 & 0.409 & 0.507 & 0.412 & 0.509 & 0.0008 & 0.0010 &  & -- & -- \\ \midrule
W+CPP & 1 & 0.569 & 0.709 & 0.429 & 0.533 & 0.0570 & 0.0726 & $\lambda$ & 5.3780 & 20.6583 \\
 & 10 & 0.423 & 0.532 & 0.315 & 0.391 & 0.0174 & 0.0217 &  & 1.1637 & 1.4508 \\
 & 100 & 0.397 & 0.502 & 0.303 & 0.378 & 0.0055 & 0.0068 &  & 0.3648 & 0.4578 \\
 & 300 & 0.407 & 0.514 & 0.298 & 0.372 & 0.0033 & 0.0041 &  & 0.2139 & 0.2682 \\ \midrule
VG & 1 & 0.577 & 0.720 & 0.105 & 0.136 & 0.0488 & 0.0634 & $\nu$ & 0.0076 & 0.0093 \\
 & 10 & 0.408 & 0.516 & 0.079 & 0.100 & 0.0165 & 0.0208 &  & 0.0039 & 0.0052 \\
 & 100 & 0.401 & 0.500 & 0.073 & 0.091 & 0.0051 & 0.0064 &  & 0.0015 & 0.0020 \\
 & 300 & 0.397 & 0.491 & 0.073 & 0.094 & 0.0029 & 0.0037 &  & 0.0009 & 0.0011 \\ \midrule
Stable & 1 & 0.797 & 1.869 & 0.264 & 0.329 & 0.0912 & 0.1289 & $\alpha$ & 0.0682 & 0.0851 \\
 & 10 & 0.833 & 3.134 & 0.179 & 0.225 & 0.0260 & 0.0327 &  & 0.0206 & 0.0259 \\
 & 100 & 0.859 & 2.127 & 0.183 & 0.228 & 0.0086 & 0.0108 &  & 0.0071 & 0.0088 \\
 & 300 & 0.886 & 2.082 & 0.180 & 0.221 & 0.0048 & 0.0061 &  & 0.0039 & 0.0050 \\ \midrule
fBm ($H=0.3$) & 1 & 0.540 & 0.678 & 0.591 & 0.756 & 0.1144 & 0.1588 & $H$ & 0.0368 & 0.0459 \\
 & 10 & 0.344 & 0.433 & 0.391 & 0.489 & 0.0328 & 0.0414 &  & 0.0113 & 0.0142 \\
 & 100 & 0.301 & 0.379 & 0.347 & 0.439 & 0.0105 & 0.0131 &  & 0.0037 & 0.0046 \\
 & 300 & 0.291 & 0.363 & 0.337 & 0.420 & 0.0057 & 0.0072 &  & 0.0020 & 0.0025 \\ \midrule
fBm ($H=0.7$) & 1 & 0.543 & 0.679 & 0.542 & 0.679 & 0.0799 & 0.0972 & $H$ & 0.0266 & 0.0333 \\
 & 10 & 0.568 & 0.722 & 0.567 & 0.720 & 0.0288 & 0.0361 &  & 0.0087 & 0.0109 \\
 & 100 & 0.680 & 0.848 & 0.679 & 0.847 & 0.0093 & 0.0117 &  & 0.0027 & 0.0034 \\
 & 300 & 0.655 & 0.834 & 0.654 & 0.833 & 0.0058 & 0.0072 &  & 0.0017 & 0.0021 \\
\bottomrule
\end{tabular}
\end{table}

\subsection{Effects of multiscale sampling frequency on parameter estimation}

This subsection examines whether the sampling frequency influences the quality of parameter estimation. To this end, we consider the PLT model \eqref{eq:general} and fix 
$\sigma=0.5$ and $T-t_0=10$. We vary the time increment size $h\in\{1,7,30\}/365$ (daily, weekly and monthly) and analyze the noises Wiener, W+CPP, VG, stable, and fBm ($H=0.3$ and $H=0.7$).

\begin{figure}[H]
\centering
\includegraphics[width=1.0\linewidth]{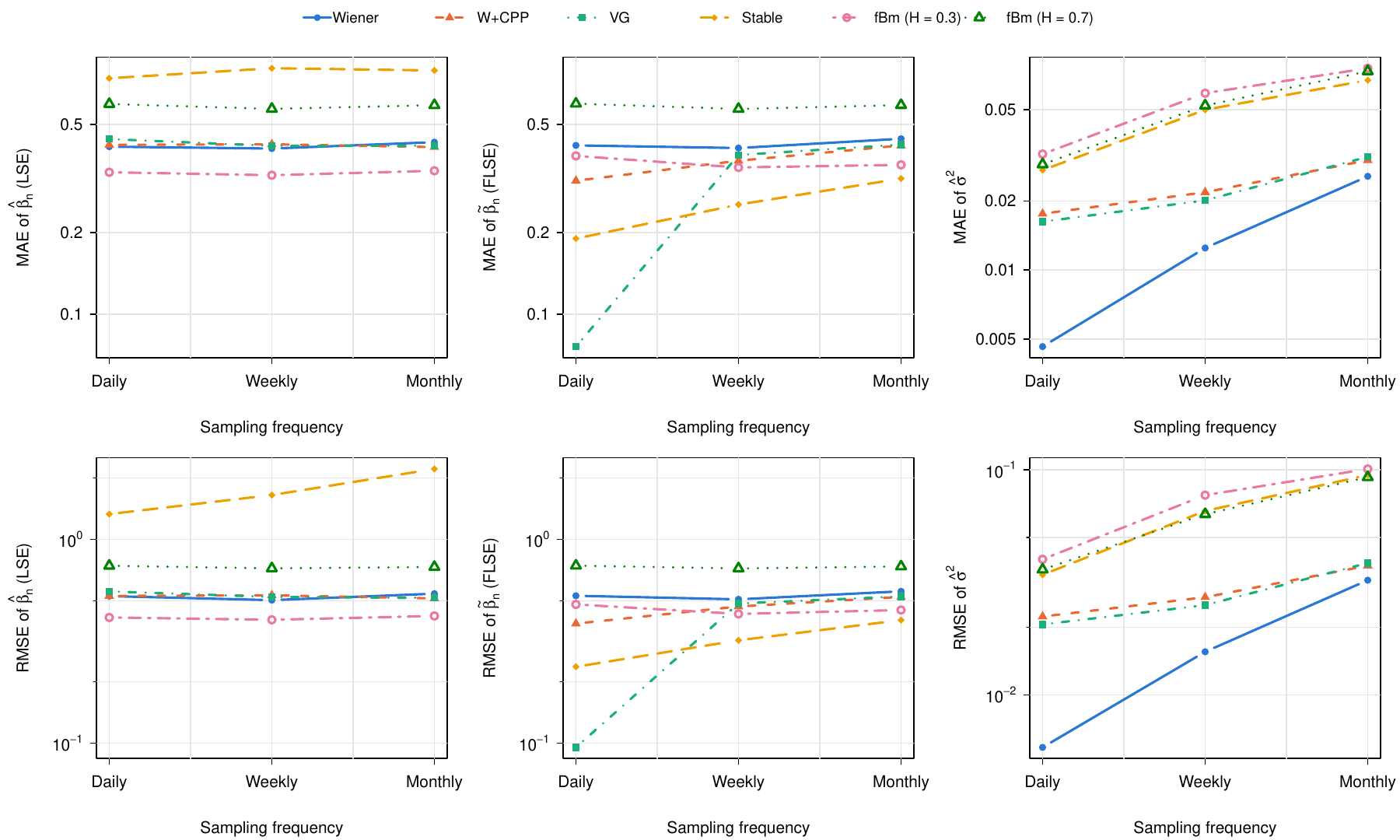}
\caption{Sampling frequency MAE (top) and RMSE (bottom) of the LSE $\hat\beta_n$ (left), the FLSE $\tilde\beta_n$ (center) and $\hat\sigma^2_n$ (right).}
\label{fig:mc3}
\end{figure}

Figure~\ref{fig:mc3} presents the MAE and RMSE from the Monte Carlo simulations. In general, weekly and monthly sampling frequencies do not affect the quality of daily estimation based on $\hat\beta$ and $\widetilde\beta$. Conversely, the poorest estimates arise from $\hat\sigma^2_n$ when data observations are more widely spaced.

\subsection{Evaluating the estimates of the PLT model with power-law volatility}\label{sec:sim_plv}

We finish this section with a simulation study on parameter estimation for the PLT model with power-law volatility \eqref{eq:plv}. This model will be the most relevant for real-data applications in the following section.

For the simulations, we consider $\sigma_0=0.5$. As stated in Theorem~\ref{prop:plv}, under the Wiener noise, the estimator $\hat\beta_n(\gamma)$ of the power parameter $\beta$ is consistent if and only if $\gamma\geq1/2$. 
Table~\ref{tab:plv} allows us to analyze the behavior of the estimator even under more general driven noises. 
Since  $\sigma_0^2$ and the noise parameter $\vartheta$ were successfully estimated in the previous subsections, we omit their evaluation here and focus solely on the $\beta$ estimates.

The $\beta$ estimates from \eqref{eq:wlse} converge to zero in terms of MAE and RMSE across all noise families. For noises with a jump component, the filtered WLSE version of the $\beta$ estimator proved more precise. The second-step estimator of $\gamma$ also yields good asymptotic results. These evaluations support the application of the PLT model with power-law volatility to real-world scenarios, such as modeling daily Bitcoin data.

\begin{table}[H]
\centering\small
\caption{MAE and RMSE of the QMLE $\hat\gamma_n$, WLSE $\hat\beta_n(\hat\gamma_n)$ and FLSE $\tilde\beta_n(\hat\gamma_n)$, based on $M=1{,}000$ Monte Carlo simulations of the PLT model with power-law volatility \eqref{eq:plv} with $\gamma=1$, $\sigma=0.5$ and $h=1/365$.}
\label{tab:plv}
\begin{tabular}{lccccccc}
\toprule
 & & \multicolumn{2}{c}{$\hat\gamma_n$} & \multicolumn{2}{c}{$\hat\beta_n(\hat\gamma_n)$} & \multicolumn{2}{c}{$\widetilde\beta_n(\hat\gamma_n)$} \\
\cmidrule(lr){3-4} \cmidrule(lr){5-6} \cmidrule(lr){7-8} 
Noise & $T-t_0$ & MAE & RMSE & MAE & RMSE & MAE & RMSE \\
\midrule
Wiener & 10 & 0.0158 & 0.0200 & 0.130 & 0.162 & 0.132 & 0.165 \\
 & 100 & 0.0035 & 0.0043 & 0.040 & 0.051 & 0.041 & 0.051 \\
 & 300 & 0.0017 & 0.0022 & 0.023 & 0.029 &  0.023 & 0.029  \\ \midrule
W+CPP & 10 & 0.0587 & 0.0735 & 0.123 & 0.157 &  0.094 & 0.118  \\
 & 100 & 0.0123 & 0.0154 & 0.040 & 0.050 &  0.029 & 0.036  \\
 & 300 & 0.0068 & 0.0086 & 0.023 & 0.028 &  0.016 & 0.021  \\ \midrule
VG & 10 & 0.0525 & 0.0660 & 0.124 & 0.155 &  0.023 & 0.029  \\
 & 100 & 0.0118 & 0.0150 & 0.041 & 0.052  & 0.007 & 0.009  \\
 & 300 & 0.0066 & 0.0081 & 0.023 & 0.028 &  0.004 & 0.005  \\ \midrule
Stable & 10 & 0.3909 & 0.5458 & 0.166 & 0.224 & 0.065 & 0.087  \\
 & 100 & 0.2497 & 0.3626 & 0.077 & 0.106 & 0.019 & 0.025 \\
 & 300 & 0.2486 & 0.3640 & 0.056 & 0.086 & 0.012 & 0.016 \\ \midrule
fBm ($H=0.3$) & 10 & 0.0164 & 0.0208 & 0.082 & 0.101  & 0.094 & 0.118  \\
 & 100 & 0.0036 & 0.0046 & 0.016 & 0.021 & 0.023 & 0.029 \\
 & 300 & 0.0020 & 0.0024 & 0.008 & 0.009 & 0.012 & 0.016 \\ \midrule
fBm ($H=0.7$) & 10 & 0.0187 & 0.0239 & 0.202 & 0.252 & 0.202 & 0.252 \\
 & 100 & 0.0043 & 0.0053 & 0.099 & 0.124 & 0.099 & 0.124 \\
 & 300 & 0.0024 & 0.0030 & 0.071 & 0.090 & 0.071 & 0.090 \\
\bottomrule
\end{tabular}
\end{table}

\newpage
\section{Applications to Bitcoin daily prices}\label{sec:application_real_data}

In this section, we present the modeling of Bitcoin daily prices. We designed all applications to demonstrate the full capabilities of the estimation approach developed here.

The series contains the daily closing prices of Bitcoin (BTC) in USD from 2010-07-13 to 2026-10-04 with $n=5{,}928$ consecutive days. We measured the time in years since the genesis block in 2009-01-03. Then, we consider $t_0 = 1.52$ and $h=1/365.25$.
We divided the dataset into two subsets: a training set ($n=5{,}863$ days) for parameter estimation and a validation set ($ n = 65$ days) for Monte Carlo forecasting and residual analysis to assess the model's goodness-of-fit.

All analyses presented here were conducted using the PLT models \ref{eq:log-plt}, \eqref{eq:general}, and \eqref{eq:plv}, comparing them under different noise structures and against the benchmark geometric Brownian motion (GBM) with the same power-law volatility, that is, the SDE is given by
$$dP(t)=\Bigl[\mu+\tfrac{1}{2}\sigma^2(t)\Bigr]P(t)\,dt+\sigma(t)P(t)\,dB(t),\qquad
\sigma(t)=\sigma_0\Bigl(\frac{t}{t_0}\Bigr)^{-\gamma},\quad t>t_0.$$

\subsection{Comparing PLT models versus benchmark GBM and noise selection}\label{sec:app_vol}

\begin{figure}[H]
\centering
\includegraphics[width=0.62\linewidth]{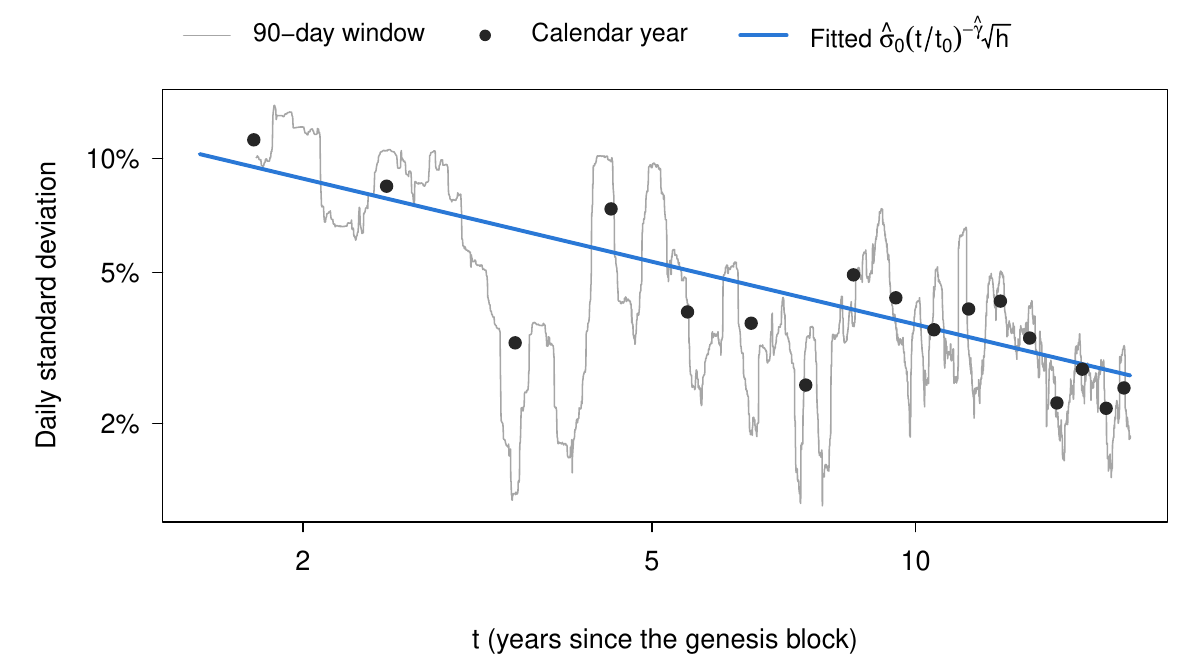}
\caption{Standard deviation of the daily log-price increments in the training window, per calendar year and in 90-day windows, and fitted power-law volatility}
\label{fig:app_vol}
\end{figure}

The standard deviation of the daily log-price increments in the training window (Figure \ref{fig:app_vol}) indicates that the volatility of Bitcoin’s daily increments decays approximately as a power of $t$, with alternating periods of high and low volatility around the trend. We begin our modeling by considering the PLT with power-law volatility \eqref{eq:plv} as a natural candidate for daily log-prices.

The QMLE \eqref{eq:wlse2} gives $\hat\gamma_n=0.55$, with robust standard error 0.063 and 95\% confidence interval (CI) $[0.426, 0.674]$. The Wald test (statistic 8.7) rejects the null hypothesis of constant volatility ($\gamma=0$). The WLSE \eqref{eq:wlse} gives $\hat\beta_n=5.601$. 
Since $\hat\gamma_n$ is close to $1/2$, the information on $\beta$ grows very slowly with $T$, as discussed in the previous sections.
The FLSE estimate $\widetilde\beta_n = 6.865$, filtering 7.1\% of the increments. 

\begin{figure}[H]
\centering
\includegraphics[width=1.0\linewidth]{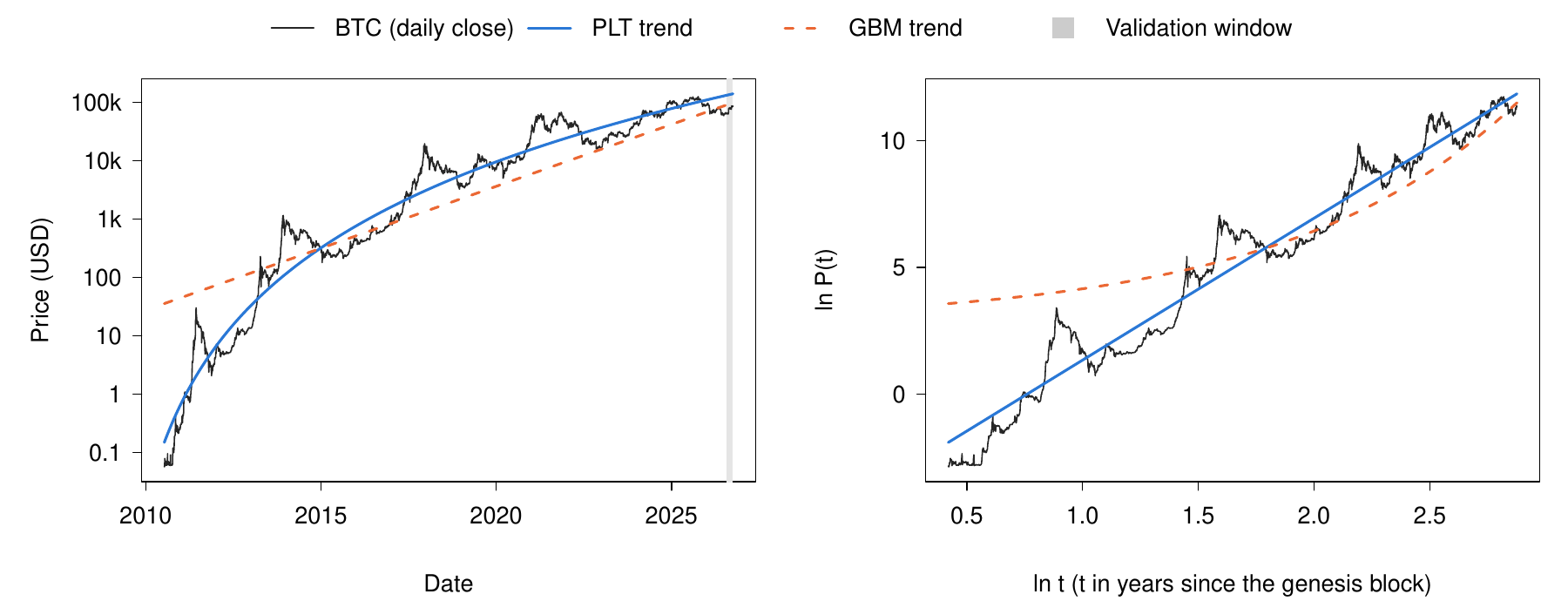}
\caption{Bitcoin daily prices model fitting using the PLT and GBM trends, in the date--price (left) and log--log (right) scales. The gray area is the validation window.}
\label{fig:app_data}
\end{figure}

Figure~\ref{fig:app_data} shows that the BTC follows the power-law trend (left panel) and a linear trend in the log--log scale (right panel). In contrast, the GBM trend is convex in this scale, overestimating prices before 2014 and underestimating them in most of the subsequent years.
This is reflected in the one-year-ahead predictions, with prediction error measures of MAE 1.088 for PLT and 1.280 for GBM.

To ensure a better model fit, noise selection was conducted through $M=1{,}000$ Monte-Carlo predictions and the corresponding error measures. Denote $X(t)=\ln P(t)$ and the $m$th Monte Carlo prediction $\hat X^{(m)}(t)$. For $k$-day-ahead forecasts, we consider the error measures
$$\operatorname{MAE}(k)=\frac{1}{k}\sum_{j=1}^{k}\frac{1}{M}\sum_{m=1}^{M}\left|\hat X^{(m)}(t_j)-X(t_j)\right|~~\text{and}~~ \operatorname{RMSE}(k)=\frac{1}{k}\sum_{j=1}^{k}\left[\frac{1}{M}\sum_{m=1}^{M}\left(\hat X^{(m)}(t_j)-X(t_j)\right)^2\right]^{1/2}.$$

Considering the previously discussed Wiener, W+CPP, VG, Stable, and fBm noises, we compare their forecasting performance for both one-day-ahead and multi-step horizons. The multi-step forecasts are generated from predictions starting on 2026-07-31 up to $k\in\{1,\cdots,65\}$ days ahead, without using any validation information in the new starting horizon. The MAE and RMSE are reported as the averages across all steps.
Table~\ref{tab:app_fit} presents the parameter estimates, the error measures for one- and multi-step forecasts, and the percentage of real-value observations lying outside the 95\% prediction bands. The results indicate frequent jumps and heavy tails, with PLT-VG being the most precise for one-step forecasts, while PLT-fBm was selected for multi-step predictions.

\begin{table}[H]
\centering 
\caption{Noise selection, estimates on the training window and validation metrics of the PLT and GBM models, all with power-law volatility.}
\label{tab:app_fit}
\resizebox{\textwidth}{!}{
\begin{tabular}{lcccccccccc}
\toprule
 & & & & & \multicolumn{3}{c}{One-step} & \multicolumn{3}{c}{Multi-step} \\
\cmidrule(lr){6-8} \cmidrule(lr){9-11}
Model & Drift & $\hat\sigma_0$ & $\hat\gamma$ & Noise parameters & MAE & RMSE & Out (\%) & MAE & RMSE & Out (\%) \\
\midrule
GBM & $\hat\mu=0.489$ & 1.969 & 0.551 & -- & 0.0266 & 0.0323 & 7.7 & 0.1666 & 0.2022 & 0.0 \\
PLT-Wiener & $\hat\beta=5.601$ & 1.963 & 0.550 & -- & 0.0265 & 0.0322 & 6.2 & 0.1781 & 0.2131 & 0.0 \\
PLT-W+CPP & $\tilde\beta=6.865$ & 1.964 & 0.550 & $\hat\lambda=159.1$, $\hat\sigma_c=0.817$, $\hat\sigma_J=0.157$ & 0.0252 & 0.0343 & 3.1 & 0.1829 & 0.2208 & 0.0 \\
\textbf{PLT-VG} & $\tilde\beta=6.865$ & 1.964 & 0.550 & $\hat\nu=0.0106$ & \textbf{0.0227} & \textbf{0.0321} & \textbf{6.2} & 0.1723 & 0.2087 & 0.0 \\
PLT-Stable & $\tilde\beta=6.865$ & 4.700 & 0.550 & $\hat\alpha=1.361$ & 0.0292 & 0.0896 & 3.1 & 0.2855 & 0.5291 & 0.0 \\
PLT-fBm & $\hat\beta=5.601$ & 2.318 & 0.550 & $\hat H=0.528$ & 0.0267 & 0.0324 & 6.2 & \textbf{0.1939} & \textbf{0.2333} & \textbf{0.0} \\
PLT-VG ($\gamma=0$) & $\tilde\beta=4.450$ & 0.895 & 0 & $\hat\nu=0.0214$ & 0.0288 & 0.0508 & 0.0 & 0.2388 & 0.3021 & 0.0 \\
\bottomrule
\end{tabular}
}
\end{table}

\subsection{Out-of-sample performance and forecasts}\label{sec:oos}

In addition to the model selection presented in Table~\ref{tab:app_fit}, we now provide a graphical verification of the selected models, together with a residual analysis.  

\begin{figure}[H]
\centering
\includegraphics[width=1.0\linewidth]{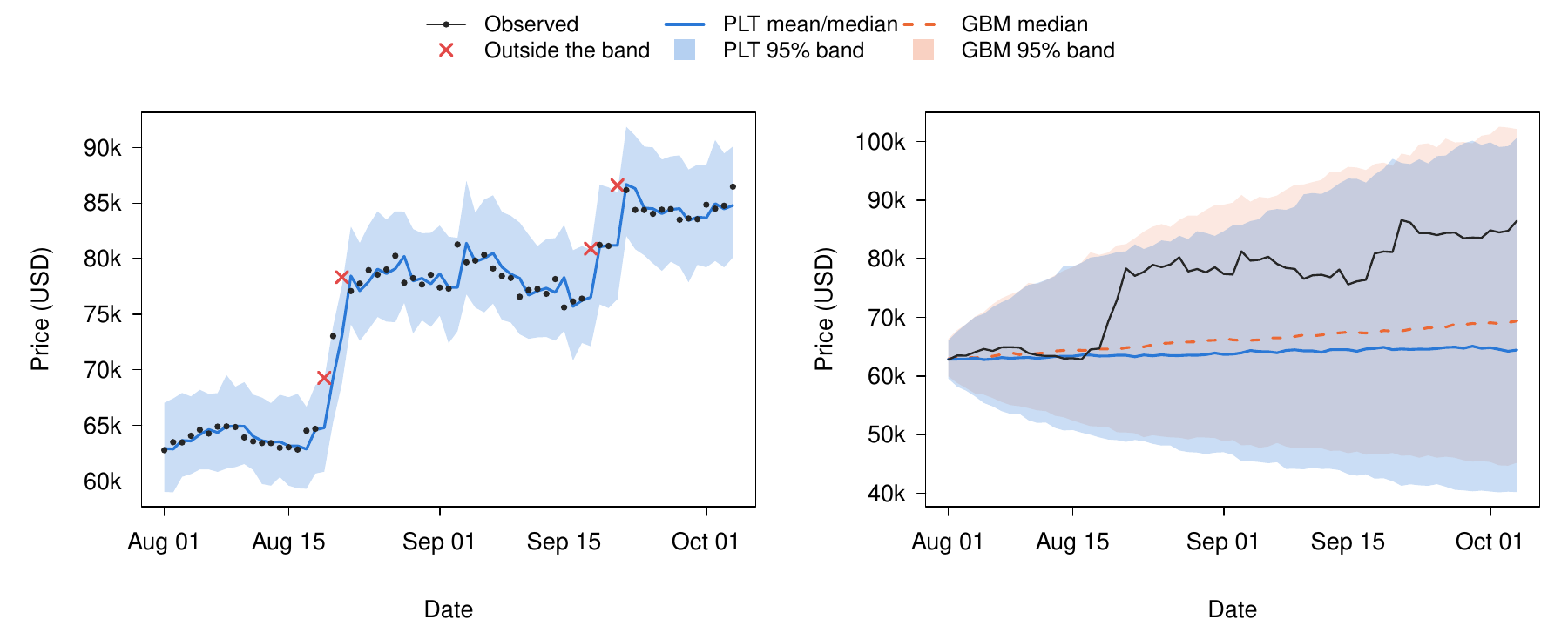}
\caption{One-step-ahead fit of the PLT-VG model with a 95\% Monte Carlo band (left), and multi-step forecasts of PLT-fBm and GBM from the end of the training window (right). The red crosses in the left panel are the prices outside the band.}
\label{fig:app_val}
\end{figure}

Figure~\ref{fig:app_val} shows the one-step fit of PLT-VG and the multi-step forecasts of PLT-fBm and GBM. For the one-step, 95\% Monte Carlo CIs leave out 6.2\% of the validation prices. No validation price falls outside the multi-step bands, which produce a graph similar to the law of the iterated logarithm for a Wiener process.
Table~\ref{tab:app_cov} reports the coverage of the 95\% Monte Carlo CIs. For the one- and multi-step forecasts, we present the percentage of out-of-band observations (Out) and the mean width of the bands (Width, computed as the Monte Carlo quantile difference $q_{0.975}-q_{0.025}$).

\begin{table}[H]
\centering\small
\caption{Percentage of observations outside (Out) the  95\% Monte Carlo bands and mean width in the coverage for one- and multi-step-ahead.}
\label{tab:app_cov}
\begin{tabular}{lcccc}
\toprule
 & \multicolumn{2}{c}{One-step} & \multicolumn{2}{c}{Multi-steps} \\
\cmidrule(lr){2-3} \cmidrule(lr){4-5}
Model & Out (\%) & Width & Out (\%) & Width  \\
\midrule
GBM & 6.9 & 0.128 &  9.1 & 1.029  \\
PLT-Wiener & 6.8 & 0.128  & 8.8 & 1.032  \\
PLT-W+CPP & 3.6 & 0.167  & 6.8 & 1.130  \\
PLT-VG & 5.4 & 0.144  & 8.4 & 1.041  \\
PLT-Stable & 3.5 & 0.168  & 0.1 & 4.910  \\
PLT-fBm & 6.9 & 0.128  & 6.2 & 1.187 \\
PLT-VG ($\gamma=0$) & 1.2 & 0.231  & 0.8 & 1.727  \\
\bottomrule
\end{tabular}
\end{table}

\begin{figure}[H]
\centering
\includegraphics[width=0.8
\linewidth]{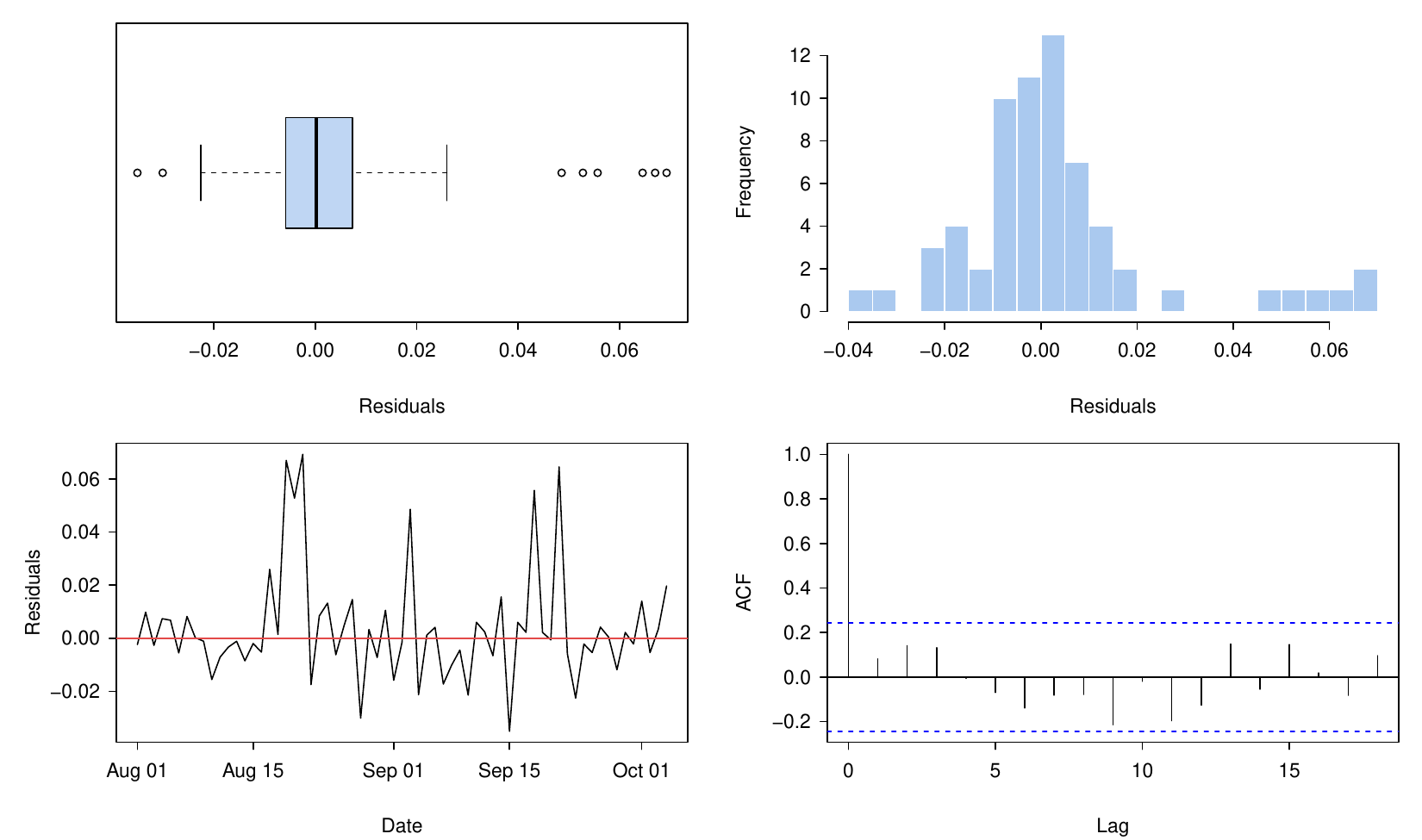}
\caption{Boxplot (top left), histogram (top right), time plot (bottom left), and autocorrelation function (bottom right) of the one-step residuals of PLT-VG model.}
\label{fig:app_res}
\end{figure}

To evaluate the quality of the fit, for each time $t_j$ in the validation window, the log-price $X(t_j)$ and the fitted process $\hat{X}(t_j)$ are compared. The residuals, calculated as $X(t_j)-\hat{X}(t_j)$, are analyzed and presented in Table \ref{tab:app_res} and Figure \ref{fig:app_res}, which show the general behavior of the residues as right-skewed, centered on zero and with no significant autocorrelation (Ljung--Box test with 10 lags, $\text{p-value}=0.50$).

\begin{table}[H]
\centering\small
\caption{Summary statistics of the one-step residuals of the PLT-VG model.}
\label{tab:app_res}
\begin{tabular}{rrrrrrrrrr}
\toprule
Min. & 1st Qu. & Median & Mean & 3rd Qu. & Max. & Sd & CS & CK & $k$ \\
\midrule
$-$0.0351 & $-$0.0058 & 0.0002 & 0.0038 & 0.0073 & 0.0693 & 0.0211 & 1.5305 & 2.6332 & 65 \\
\bottomrule
\end{tabular}
\end{table}

\begin{figure}[H]
\centering
\includegraphics[width=1.0\linewidth]{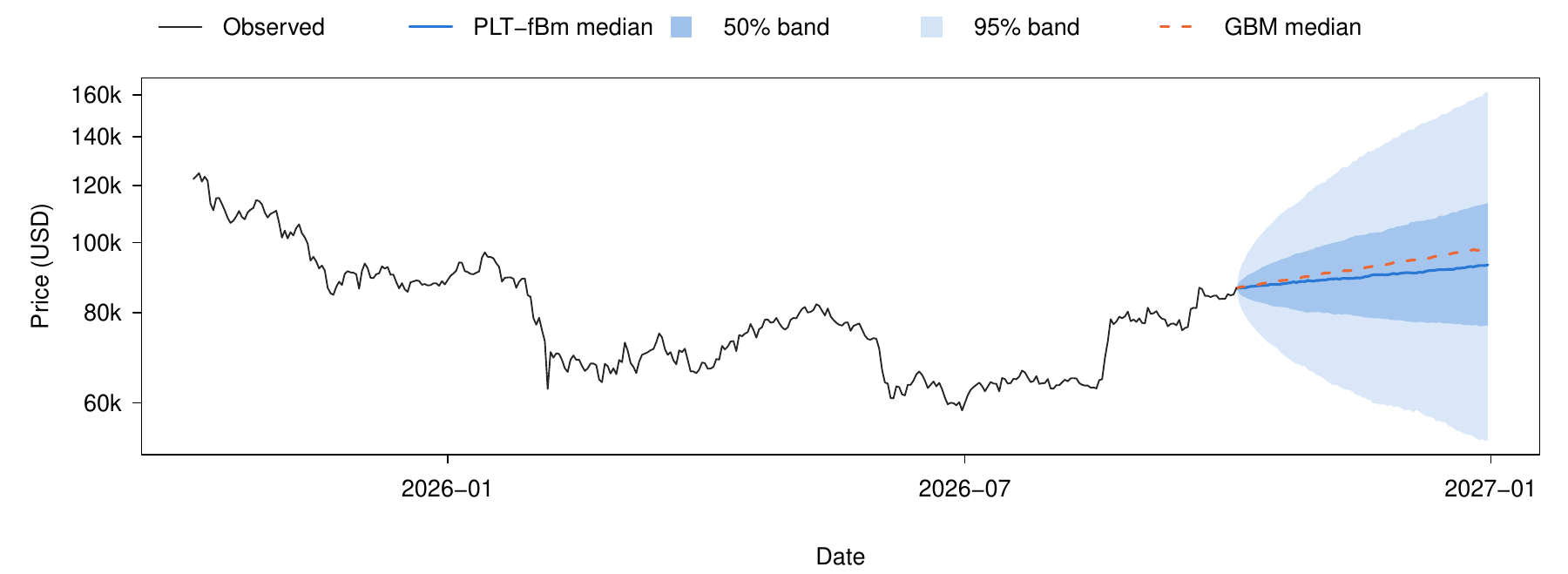}
\caption{Monte Carlo multi-step forecast for unobserved days.}
\label{fig:app_fcst}
\end{figure}

Re-estimating the PLT-fBm parameters to the whole series gives $\hat\gamma_n=0.554$ and $\hat{\beta}_n=5.717$. Figure~\ref{fig:app_fcst} shows the Monte Carlo forecast ($M=5{,}000$) up to 2026-12-31: starting from US\$86480 on 2026-10-04, the median is US\$93146, with 95\% band $[53255, 161687]$.

\section{Conclusions}\label{sec:6}

In this paper, we introduced power-law trend SDEs for Bitcoin modeling. Thus, the new class of SDEs was built based on the log–log empirical properties of Bitcoin. A final model with power-law volatility was further defined and performed in practical settings.

The present work evaluated the advantages of considering noises more general than a simple Wiener process.
Lévy processes were used for jump modeling, while fBm was employed to capture long-term behavior.

Least-squares-based estimators were proposed and adapted for each model discussed. The asymptotic behavior was studied theoretically, while Monte Carlo simulations attested to the performance of the analytical closed-form expressions hereby derived.

The framework explored in this work can serve as a starting point for studying models with power-law trends, such as new parameter estimation procedures, analytical properties of the model, and the derivation of option pricing formulas.

\small

\bibliographystyle{unsrt}
\bibliography{sample}

@article{carvalho2025multiscale,
  title={Multiscale Stochastic Models for Bitcoin: Fractional Brownian Motion and Duration-Based Approaches},
  author={Carvalho, Arthur Rodrigues Pereira de and Quintino, Felipe and Saulo, Helton and Ozelim, Luan CSM and Fonseca, Tiago A da and Rathie, Pushpa N},
  journal={FinTech},
  volume={4},
  number={3},
  pages={51},
  year={2025},
  doi={10.3390/fintech4030051},
  publisher={MDPI}
}

@book{mikosch1998elementary,
  title={Elementary stochastic calculus with finance in view},
  author={Mikosch, Thomas},
  year={1998},
  publisher={World scientific}
}

@article{sornette2025transaction,
  title={Transaction flows and holding time scaling laws of bitcoin},
  author={Sornette, Didier and Zhang, Yu},
  journal={Physica A: Statistical Mechanics and its Applications},
  volume={658},
  pages={130294},
  year={2025},
  publisher={Elsevier}
}

@article{li2025machine,
  title={Machine learning, memory and efficiency in cryptocurrency markets},
  author={Li, Shuyue and Yarovaya, Larisa and Mishra, Tapas},
  journal={Journal of International Financial Markets, Institutions and Money},
  volume={105},
  pages={102210},
  year={2025},
  publisher={Elsevier}
}

@article{shu20212021,
  title={The 2021 bitcoin bubbles and crashes—detection and classification},
  author={Shu, Min and Song, Ruiqiang and Zhu, Wei},
  journal={Stats},
  volume={4},
  number={4},
  pages={950--970},
  year={2021},
  publisher={MDPI}
}

@article{huang2023time,
  title={{Time-varying dependence between Bitcoin and green financial assets: A comparison between pre-and post-COVID-19 periods}},
  author={Huang, Yingying and Duan, Kun and Urquhart, Andrew},
  journal={Journal of International Financial Markets, Institutions and Money},
  volume={82},
  pages={101687},
  year={2023},
  publisher={Elsevier}
}

@article{yaya2022modelling,
  title={{Modelling cryptocurrency high--low prices using fractional cointegrating VAR}},
author = {Yaya, OlaOluwa S. and Vo, Xuan Vinh and Ogbonna, Ahamuefula E. and Adewuyi, Adeolu O.},
  journal={International Journal of Finance \& Economics},
  volume={27},
  number={1},
  pages={489--505},
  year={2022},
  publisher={Wiley Online Library}
}

@article{wu2020long,
  title={Long memory and efficiency of Bitcoin under heavy tails},
  author={Wu, Liang and Chen, Shujuan},
  journal={Applied Economics},
  volume={52},
  number={48},
  pages={5298--5309},
  year={2020},
  publisher={Taylor \& Francis}
}

@article{wheatley2019bitcoin,
  title={{Are Bitcoin bubbles predictable? Combining a generalized Metcalfe’s law and the log-periodic power law singularity model}},
  author={Wheatley, Spencer and Sornette, Didier and Huber, Tobias and Reppen, Max and Gantner, Robert N},
  journal={Royal Society Open Science},
  volume={6},
  number={6},
  year={2019},
  publisher={The Royal Society}
}

@article{wei2018liquidity,
  title={Liquidity and market efficiency in cryptocurrencies},
  author={Wei, Wang Chun},
  journal={Economics Letters},
  volume={168},
  pages={21--24},
  year={2018},
  publisher={Elsevier}
}

@article{ciaian2018virtual,
  title={Virtual relationships: Short-and long-run evidence from BitCoin and altcoin markets},
  author={Ciaian, Pavel and Rajcaniova, Miroslava and others},
  journal={Journal of International Financial Markets, Institutions and Money},
  volume={52},
  pages={173--195},
  year={2018},
  publisher={Elsevier}
}

@article{stavroyiannis2019high,
  title={The high frequency multifractal properties of Bitcoin},
  author={Stavroyiannis, Stavros and Babalos, Vassilios and Bekiros, Stelios and Lahmiri, Salim and Uddin, Gazi Salah},
  journal={Physica A: Statistical Mechanics and its Applications},
  volume={520},
  pages={62--71},
  year={2019},
  publisher={Elsevier}
}

@article{beguvsic2018scaling,
  title={Scaling properties of extreme price fluctuations in Bitcoin markets},
  author={Begu{\v{s}}i{\'c}, Stjepan and Kostanj{\v{c}}ar, Zvonko and Stanley, H Eugene and Podobnik, Boris},
  journal={Physica A: Statistical Mechanics and its Applications},
  volume={510},
  pages={400--406},
  year={2018},
  publisher={Elsevier}
}

@article{sensoy2019inefficiency,
  title={The inefficiency of Bitcoin revisited: A high-frequency analysis with alternative currencies},
  author={Sensoy, Ahmet},
  journal={Finance Research Letters},
  volume={28},
  pages={68--73},
  year={2019},
  publisher={Elsevier}
}

@article{nadarajah2017inefficiency,
  title={On the inefficiency of Bitcoin},
  author={Nadarajah, Saralees and Chu, Jeffrey},
  journal={Economics Letters},
  volume={150},
  pages={6--9},
  year={2017},
  publisher={Elsevier}
}

@article{bariviera2017inefficiency,
  title={The inefficiency of Bitcoin revisited: A dynamic approach},
  author={Bariviera, Aurelio F},
  journal={Economics Letters},
  volume={161},
  pages={1--4},
  year={2017},
  publisher={Elsevier}
}

@article{urquhart2016inefficiency,
  title={The inefficiency of Bitcoin},
  author={Urquhart, Andrew},
  journal={Economics Letters},
  volume={148},
  pages={80--82},
  year={2016},
  publisher={Elsevier}
}

@book{antonopoulos2023mastering,
  title={Mastering Bitcoin: Programming the open blockchain},
  author={Antonopoulos, Andreas M and Harding, David A},
  year={2023},
  publisher={"O'Reilly Media, Inc."}
}

@article{nakamoto2008bitcoin,
  title={Bitcoin: A Peer-to-Peer Electronic Cash System},
  author={Nakamoto, Satoshi},
  year={2008},
  url={https://bitcoin.org/bitcoin.pdf}
}

@book{antonopoulos2014mastering,
  title={Mastering Bitcoin: Unlocking Digital Cryptocurrencies},
  author={Antonopoulos, Andreas M.},
  year={2014},
  publisher={O'Reilly Media},
  pages = {89--101},
  edition = {1st}
}

@article{garnier2019chaos,
title = {Chaos and order in the bitcoin market},
author = {Josselin Garnier and Knut Solna},
journal = {Physica A: Statistical Mechanics and its Applications},
volume = {524},
pages = {708--721},
year = {2019},
url = {https://doi.org/10.1016/j.physa.2019.04.164}
}

@article{dimpfl2020bitcoin,
  title={Bitcoin Price Risk—A Durations Perspective},
  author={Dimpfl, Thomas and Odelli, Stefania},
  journal={Journal of Risk and Financial Management},
  volume={13},
  number={7},
  pages={157},
  year={2020},
  publisher={MDPI},
  url ={https://www.mdpi.com/1911-8074/13/7/157}
}

@article{Mancini09,
author = {Mancini, C.},
title = {Non-parametric Threshold Estimation for Models with Stochastic Diffusion Coefficient and Jumps},
journal = {Scandinavian Journal of Statistics},
volume = {36},
number = {2},
pages = {270-296},
doi = {https://doi.org/10.1111/j.1467-9469.2008.00622.x},
year = {2009}
}

@article{Rejeb2023,
  title = {Tracing Knowledge Diffusion Trajectories in Scholarly Bitcoin Research: Co-Word and Main Path Analyses},
  volume = {16},
  ISSN = {1911-8074},
  url = {http://dx.doi.org/10.3390/jrfm16080355},
  DOI = {10.3390/jrfm16080355},
  number = {8},
  journal = {Journal of Risk and Financial Management},
  publisher = {MDPI AG},
  author = {Rejeb,  Abderahman and Rejeb,  Karim and Alnabulsi,  Khalil and Zailani,  Suhaiza},
  year = {2023},
  pages = {355}
}

@article{MeredizSol2019,
  title = {A bibliometric analysis of bitcoin scientific production},
  volume = {50},
  ISSN = {0275-5319},
  url = {http://dx.doi.org/10.1016/j.ribaf.2019.06.008},
  DOI = {10.1016/j.ribaf.2019.06.008},
  journal = {Research in International Business and Finance},
  publisher = {Elsevier BV},
  author = {Merediz-Solà,  Ignasi and Bariviera,  Aurelio F.},
  year = {2019},
  pages = {294–305}
}

@article{quintino2026asymptotics,
  title={Asymptotics for drift estimation of a {L}{\'e}vy-driven generalized {L}angevin equation sampled at high-frequency},
  author={Quintino, Felipe S. and Medino, Ary V. and Dorea, Chang C. Y.},
  journal={Journal of Applied Statistics},
  year={2026},
  doi={10.1080/02664763.2026.2649343}
}

\end{document}